\documentclass[aps,prb,twocolumn,superscriptaddress,floatfix,showpacs]{revtex4}
\usepackage{CJK}
\bibpunct{[}{]}{,}{n}{}{}
\usepackage{graphicx}
\usepackage{dcolumn}
\usepackage{float}
\usepackage{longtable}
\usepackage{latexsym}
\usepackage{amssymb}
\usepackage{amsfonts}
\usepackage{amsmath}
\usepackage{fancybox}
\usepackage[thinlines]{easytable}
\usepackage[binary,squaren]{SIunits}
\usepackage{natbib,graphicx,dcolumn,color,lipsum}
\usepackage{subfigure,amsmath,amssymb,verbatim,moreverb,bm,framed}
\usepackage{color}
\usepackage{colordvi}
\usepackage{ulem}
\usepackage{psfrag}

\def\be{\begin{equation}}
\def\ee{\end{equation}}
\def\ber{\begin{eqnarray}}
\def\eer{\end{eqnarray}}

\def\kv{{\bf k}}

\newcommand{\commentout}[1]{}

\DeclareMathAlphabet\mathbfcal{OMS}{cmsy}{b}{n}

\def\be{\begin{equation}}
\def\ee{\end{equation}}
\def\ber{\begin{eqnarray}}
\def\eer{\end{eqnarray}}
\def\ber*{\begin{eqnarray*}}
\def\eer*{\end{eqnarray*}}

\begin{document}

\title{Theory of the spin galvanic effect at oxide interfaces}

\author{G\"otz Seibold}
\affiliation{Institut f\"ur Theoretische Physik, BTU, Cottbus-Senftenberg, PBox 101344, 03013 Cottbus, Germany}
\author{Sergio Caprara}
\affiliation{Dipartimento di Fisica Universit\`a di Roma Sapienza, piazzale Aldo Moro 5, I-00185 Roma, Italy}
\author{Marco Grilli}
\affiliation{Dipartimento di Fisica Universit\`a di Roma Sapienza, piazzale Aldo Moro 5, I-00185 Roma, Italy}
 \author{Roberto Raimondi}
\affiliation{Dipartimento di Matematica e Fisica, Universit\`a Roma Tre,
Via della Vasca Navale 84, 00146 Rome, Italy}


\date{\today}
\begin{abstract}
The spin galvanic effect (SGE) describes the conversion of a non-equilibrium spin polarization into a transverse 
charge current. Recent experiments have demonstrated a large conversion efficiency for the two-dimensional electron gas 
formed at the interface between two insulating oxides, LaAlO$_3$ and SrTiO$_3$. Here we analyze the SGE for oxide 
interfaces within a three-band model for the Ti t$_{2g}$ orbitals which displays an interesting variety of effective
spin-orbit couplings in the individual bands that contribute differently to the spin-charge conversion. Our analytical 
approach is supplemented by a numerical treatment where we also investigate the influence of disorder and temperature, which 
turns out to be crucial to provide an appropriate description of the experimental data.
\end{abstract}

\maketitle

The spin galvanic effect (SGE) is the generation of an electrical current by a non-equilibrium spin density.
The latter may be obtained by optical or magnetic pumping. There exists also the inverse effect (ISGE) by which
the spin of free carriers can be oriented by an applied electric field. The SGE effect was first predicted \cite{Ivchenko1978} 
and then observed for Te crystals \cite{Vorobev1979}.
About a decade later, the SGE was studied theoretically \cite{Ivchenko1989,Edelstein1990,Aronov1989} in the two-dimensional 
electron gas (2DEG) in the presence of Rashba spin-orbit coupling (SOC) arising from the asymmetry of the quantum 
well \cite{Rashba84}.  The effect was later observed in quantum wells by absorption of polarized 
light \cite{GanichevPRL01,Ganichev02,Ganichev2006}.
Whereas at microscopic level the coupling between the spin polarization and the electrical current is provided by the 
SOC, the origin of the effect rests on the restricted symmetry conditions of gyrotropic media, where polar 
and axial vectors transform according to the same representation (c.f. the review by Ganichev et al., 
\cite{Ganichev2016}).
The conditions of sizable SOC and lack of inversion symmetry can be obtained also in other physical systems. Recently, indeed, 
the SGE has been observed at a silver-bismuth interface where the non-equilibrium spin polarization has been pumped by an 
adjacent ferromagnetic layer with a precessing magnetization \cite{Sanchez13}. Magnetic spin pumping was successively used to 
observe the SGE in a number of different interface systems as in ferromagnetic-topological insulators \cite{Mellnik2014,Shiomi2014} 
and ferromagnetic-oxide systems \cite{chauleau16,Lesne2016,Song2017}. In the latter case, the oxide being a heterostructure of LaAlO$_3$ (LAO) 
and SrTiO$_3$ (STO), the complex band structure originating from the Ti orbitals may lead to a richer phenomenology
\cite{hwang12,sc16}, as compared
to other systems where the standard continuum 2DEG model with spin-orbit split bands provides a good quantitative and 
qualitative understanding of the effect. Indeed, the spin-polarization induced transverse voltage $V_{SGE}$ can be modulated 
by gating the heterostructure which changes the chemical potential $\mu$, i.e. the carrier density in the 2D interface layer. 
Quite dramatically, a sign change occurs \cite{Lesne2016} in $V_{SGE}$ at a particular value of $\mu$, and has been attributed 
to a Lifshitz transition, where the $d_{xz}$, $d_{yz}$ bands originating from the Ti $t_{2g}$ orbitals start to be filled. Moreover, 
the effect strongly depends on temperature and on the number of LAO layers in the LAO/STO heterostructure \cite{Song2017}.

\begin{figure}[htb]
      \hspace*{-1cm}\includegraphics[width=14cm,clip=true]{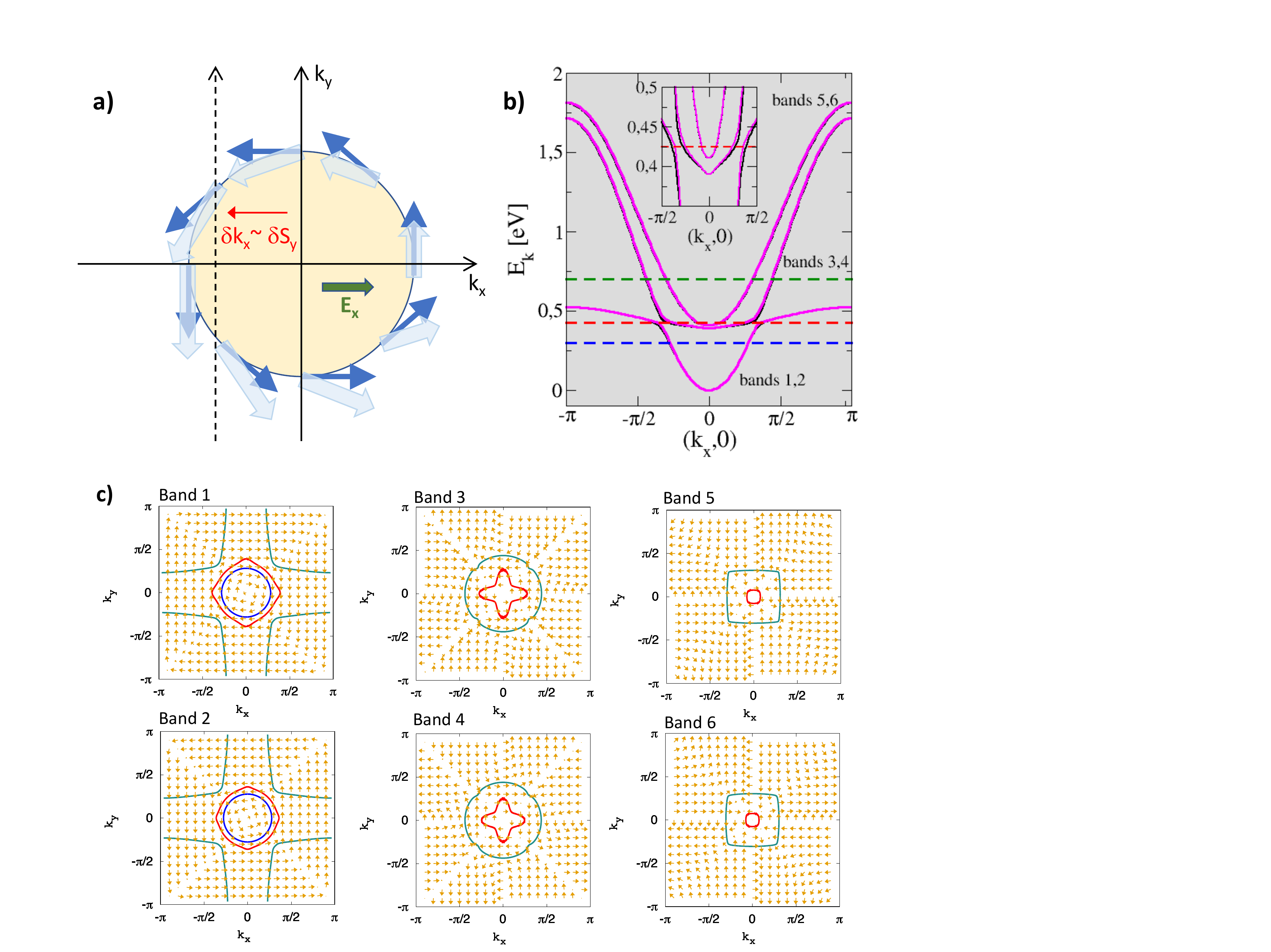}
      \caption{a) Schematic physical picture of the origin of the ISGE. b) 
       Structure of the $t_{2g}$ interface bands.
       The inset enlarges the region around the Lifshitz point where the spin-orbit splitting is large. 
       The horizontal dashed  lines in the main panel refer to the chemical potentials $\mu =0.3$\,eV (blue), $\mu =0.425$\,eV 
       (red) and $\mu =0.7$\,eV (green).
       c) The chiral spin structure for the upper band of each of the three
       pair of bands together with the Fermi surface. For the lower bands the spins point
       in the opposite direction.}
      \label{fig1}
\end{figure}

In Fig.\,\ref{fig1} a) we sketch the physical origin of the ISGE in a 2DEG with standard parabolic spectrum. In the presence 
of an applied electric field $E_x$ along the $x$ direction, the distribution function is shifted in momentum space by an amount 
$\delta k_x =-e E_x \tau$, where $e>0$ is the unit charge and $\tau$ the elastic scattering time. In the presence of Rashba 
SOC \cite{Rashba84}, $H_{RSOC}=\alpha \hbar (\tau_x k_y -\tau_y k_x)$ with $\tau_i$ denoting the Pauli matrices, each momentum 
state sees an effective magnetic field (blue thin arrows) perpendicular to the momentum, see
Fig.\,\ref{fig1} a). The presence of the electric field adds up a non-equilibrium magnetic field directed along the negative $y$
axis. As a result the magnetic field (light blue tick arrows) seen at each momentum state is no longer just opposite for $k_x$ and 
$-k_x$ states. This difference in magnetic field, which is linear in the SOC constant $\alpha$, leads to a net spin polarization 
along the $y$ axis, $S_y =\sigma^{ISGE} E_x$ \cite{Edelstein1990}, where 
$\sigma^{ISGE}=-e \alpha N_0 \tau$ denotes the inverse
spin galvanic response and 
$N_0$ is the two-dimensional density of states (DOS)  per spin of 
the parabolic band. Hence, the sign of the ISGE and, by Onsager reciprocity arguments, that of the SGE $\sigma^{SGE}$, is a direct consequence of 
the sign of the Rashba SOC. Because both the spin polarization $S_y$ and the charge current $J_x$ are 
odd under time reversal, SGE and ISGE are the same, thus sharing the same sign.

In this letter we provide the theory of the SGE and ISGE for the 2DEG at the LAO/STO interface. As it is 
customary \cite{Zhong2013,Khalsa2013,Kim2013}, the latter is described by a tight-binding model of the Ti t$_{2g}$  orbitals 
supplemented by atomic spin-orbit interactions  and an inter-orbital hopping induced by the interface asymmetry. Our 
aim is to understand what are the peculiar features of the SGE in an effectively six-band model (the $d_{xy}$, $d_{xz}$ and 
$d_{yz}$ orbitals with the additional spin degree of freedom) and to what extent it can be described in terms of  the concepts 
developed for the Rashba SOC in the 2DEG in semiconductor quantum wells. More specifically, we address the question whether 
the sign of the SGE can be related to the orientation of the internal magnetic field with respect to the momentum.
To this end we use a numerical technique, which allows us to evaluate exactly, in a finite system, the Kubo formula for the spin 
density-charge current response function \cite{ShenVR14}. Close to the $\Gamma$ point in momentum space, the tight-binding 
band structure can be approximated by a continuum model, which can be tackled analytically by means of standard 
quantum-field theory techniques normally employed in the study of disordered electron systems. This allows us to develop a 
picture of the sign behavior of the SGE as a function of the position of the chemical potential in the band structure.

The tight-binding Hamiltonian is composed of three parts
\be
H=H_0+H_{aso}+H_{I},
\label{hamiltonian}
\ee
describing the tight-binding hopping of the Ti t$_{2g}$ orbitals, the atomic SOC and the 
inter-orbital hopping, respectively.
The first term, $H_0$, is diagonal in the basis $(|xy\rangle\otimes|\sigma\rangle, \ |xz\rangle\otimes|\sigma\rangle, 
\ |yz\rangle\otimes|\sigma\rangle)$, 
$\sigma=\uparrow,\downarrow$, with spin-degenerate energies
\ber*
 \varepsilon^{xy}_k&=& -2t_1\lbrack \cos(k_x)+\cos(k_y)-2\rbrack \\
 &~&-4t_3\lbrack\cos(k_x)\cos(k_y)-1\rbrack,\label{enxy}\\
 \varepsilon^{xz}_k&=& -2(t_1+t_3)\lbrack\cos(k_x)-1\rbrack-2t_2\lbrack\cos(k_y)-1\rbrack +\Delta, \label{enxz}\\
 \varepsilon^{yz}_k&=& -2(t_1+t_3)\lbrack\cos(k_y)-1\rbrack-2t_2\lbrack\cos(k_x)-1\rbrack +\Delta,\label{enyz}
\eer*
where the energy difference $\Delta$ between $|xy\rangle$
and $|xz\rangle, |yz\rangle$ states is due to the confinement of the
2DEG near the interfacial $xy$-plane. The standard ${\bf l}\cdot {\bf s}$ coupling leads to 
the atomic SOC Hamiltonian
    \[
    %\label{eq:hatso}
      H_{aso}=\Delta_{aso}\left(\begin{array}{ccc}
        0 & -i\tau_x & i \tau_y \\
        i\tau_x & 0 & -i\tau_z \\
        -i\tau_y & i \tau_z & 0
      \end{array}
      \right).
    \]
Finally, the crystal symmetry is lifted at the interface and the orbitals get coupled through the spin independent Hamiltonian
\[
%\label{eq:ashop}
      H_{I}=\gamma\left(\begin{array}{ccc}
        0 & -2i\sin(k_y) & -2i\sin(k_x) \\
        2i\sin(k_y) & 0 & 0 \\
        2i\sin(k_x) & 0 & 0
      \end{array}
      \right)\,.
\]
In Fig.\,\ref{fig1} b) we plot the band structure of the Hamiltonian of Eq.\,(\ref{hamiltonian}), with
$t_1=0.277$\,eV, $t_2=0.031$\,eV, $t_3=0.076$\,eV, $\Delta=0.4$\,eV, $\Delta_{aso}=0.01$\,eV, 
$\gamma=0.02$\,eV, as derived in Ref.\,\cite{Zhong2013}
from projecting DFT on the $t_{2g}$ Wannier states \cite{note1}. Bands come in spin-orbit split pairs indicated by 
black and magenta curves. The pair of bands 1 and 2 originates mostly from the $d_{xy}$ orbital, whereas pairs 3,4 and 5,6 from 
the mixing of $d_{xz}$ and $d_{yz}$ due to the interface asymmetry term in the Hamiltonian. The horizontal dashed lines (red, blue 
and green) show different positions of the Fermi level in the bands. In the Rashba Hamiltonian, the eigenstates at fixed momentum 
$\kv$ form a Kramers doublet with opposite helicity, defined as the spin projection along the quantization axis, which is 
perpendicular to the momentum. In Fig.\,\ref{fig1} c) we show the vector plots of the chiral spin structure for the three pairs 
of bands. As we will show, the analysis of the behavior of the helicity of the band eigenstates provides a useful guide to 
understand the origin of the sign of the SGE obtained by a quantitative calculation.

Let us concentrate first on the lowest pair of bands (1,2). At low electron filling, the Fermi surface is circular (cf. 
the blue line in the first column of panel c in Fig.\,\ref{fig1}).  The helicity of the two spin-split bands is opposite 
and close to $\Gamma$ is given by $\Lambda\sim{\vec S_{12}}\sim s\, {\vec e_{\phi}}$, where ${\vec e_{\phi}}$ denotes
the tangential unit vector $(-\sin(\phi), \cos(\phi))$ and here and in the following $s=\mp$ refers to the lower (1,3,5) and 
upper (2,4,6) of each pair of bands, respectively. The structure therefore resembles that of the Rashba model but for an overall 
sign. Indeed, expansion near the $\Gamma$ point \cite{Kim2013,zhou2015} leads to an effective Hamiltonian 
$H_{12}=\alpha_{xy} (k_y\tau_x - k_x\tau_y)$ for the (1,2) two bands, i.e. to a Rashba Hamiltonian with a {\it negative} 
coupling constant $\alpha_{xy}=-4\gamma\Delta_{aso}/\Delta$. By increasing the filling, the Fermi surface acquires a 
distortion compatible with the crystal symmetry, which for the Rashba Hamiltonian is ruled by the $C_{4v}$ point group. 
However, the overall sign behavior of the helicity does not change.
%The linear Rashba model is special in the sense that the 
%sign of the coupling also determines the helicity.
%This is in general no longer valid for more complex SOCs as those determining
%the structure of bands (3,4) and (5,6), respectively.
For bands (3,4) the effective Hamiltonian close to $\Gamma$ reads 
$H_{34}=-\beta (k_x^2-k_y^2) (\tau_x k_y-\tau_y k_x)$ with $\beta =\gamma(t_1+t_3-t_2)/\Delta$, i.e. a Rashba-like Hamiltonian 
with a momentum-dependent coupling $\sim -\beta (k_x^2-k_y^2)$. At first sight, it seems that the factor $k_x^2-k_y^2$ spoils 
the $C_{4v}$ symmetry. This can be seen by noticing that under a $\pi/4$-rotation, Pauli matrices and momentum components transform 
as $\tau_x \rightarrow \tau_y, \tau_y \rightarrow -\tau_x$ and $k_x \rightarrow k_y, k_y \rightarrow -k_x$. Such a violation 
of symmetry does not occur, though, because the spin operator for the $S_y$ polarization, when projected onto the sector of the 
(3,4) bands, becomes $S_y =-\kappa (k_x^2-k_y^2)\tau_y /2$ with $\kappa=(t_1+t_3-t_2)/(4\Delta_{aso})$, thus restoring the 
correct symmetry at the level of physical observables.
Including the additional prefactor from the spin projection therefore yields
a helical spin structure close to
$\Gamma$ ${\vec S_{34}}\sim -s |\cos(2\phi)| {\vec e_{\phi}}\sim -|\cos(2\phi)|{\vec S_{12}}$, i.e. a helicity opposite to that
of bands 1(2). 
Finally, in the case of bands (5,6), the effective Hamiltonian at low filling reads $H_{56}=\alpha (\tau_x k_y+\tau_y k_x)$ 
with $\alpha=4\gamma\Delta_{aso}/\Delta$. Even in this case, as for the bands (3,4), the form of the Hamiltonian appears 
to violate the $C_{4v}$ symmetry and in fact is Dresselhaus-like and obeys the $D_{2d}$ symmetry. However, also in this 
case the correct spin operator in the restricted basis is $S_y=\kappa (k_x^2-k_y^2)\tau_y /2$ and restores the $C_{4v}$ symmetry. 
The helical spin structure near $\Gamma$, 
${\vec S_{56}}\sim s \cos(2\phi) \tau_z {\vec e_{\phi}}\sim -\cos(2\phi)\tau_z{\vec S_{12}}$, has a helicity which is opposite
to that of the pair (3,4).
%However, since the SOCs in $H_{56}$ and $H_{34}$ are opposite,
%one may expect that both pairs of bands may contribute with the same sign to the SGE.

As outlined above in the diffusive limit the sign of the spin galvanic response
for the Rashba model is intimately connected to the sign of the
SOC and thus to the helicity of the bands. Table \ref{tab1} summarizes the
argument by providing in the 5th row the result for the spin galvanic
response obtained in the diffusive limit in the vicinity of 
the $\Gamma$ point by using standard field-theory impurity techniques.
Clearly, the helicity $\Lambda$ for
each pair of bands determines the sign of $\sigma^{SGE}$ which
in Tab. \ref{tab1} is given in terms of the Fermi momentum $p_F$ and
the coupling constants defined above. Close to the $\Gamma$ point the bands, in the absence of SOC, are parabolic with a circular Fermi surface of radius $p_F$.
From this analysis we therefore may expect a sign change in $\sigma^{SGE}$
upon going from bands (1,2) to bands (3,4) and a further sign change
upon shifting the chemical potential into the bottom of bands (5,6).
As already mentioned, Onsager reciprocity $\sigma^{ISGE}=\sigma^{SGE}$
implies the same result for the inverse SGE.

However, while the above reasoning provides a detailed understanding
of the SGE in the individual bands the total response involves
the sum of these intraband contributions and, moreover, is expected also to
depend on scattering processes between the bands, in particular close
to the Lifshitz point.

%Let us summarize the argument by moving from left to right along the first row in panel c) of Fig.\,\ref{fig1}. In band 1, 
%states with positive $k_x$ momentum have negative helicity. Corresponding states in band 3, instead, have positive helicity. 
%Since in both cases the coupling constant is negative, the SGE is expected to change sign. Finally, the corresponding states in 
%band 5 have negative helicity, but opposite sign for the coupling constant. Hence, one expects no sign change for the SGE in 
%going from band 3 to band 5. This conclusion, based on the independent contributions of the three pairs of bands is indeed 
%confirmed by an explicit calculation of the clean limit of the SGE, as shown in Tab.\,\ref{tab1}.   
%By clean limit we mean the somewhat artificial case when there is no dissipation and the momentum relaxation time $\tau$ becomes 
%infinite. It is well known that in this limit the frequency dependent Drude longitudinal conductivity changes from a Lorentzian 
%line shape to a delta-like peak. In complete analogy, for the ISGE we define \cite{ShenVR14} the DC conductivity through
%$S_y=\sigma^{SGE} E_x$,
%which evolves to $\sigma^{SGE}=D\delta (\omega)$. The weight $D$ is shown in the 5th row of Tab.\,\ref{tab1}.

%\begingroup
%  \squeezetable
%{\small
\begin{table*}
\begin{TAB}(r,0.3cm,0.4cm)[4pt]{|c|c|c|c|}{|c|c|c|c|c|c|}
% after \\ : \hline or \cline{col1-col2} \cline{col3-col4} ...
 Bands  & $(1,2)$ &$(3,4)$ & $(5,6)$ \\ 
 $H$  & $-\alpha (\tau_x k_y-\tau_y k_x)$ &$-\beta (k_x^2-k_y^2) (\tau_x k_y-\tau_y k_x)$  &$\alpha (\tau_x k_y+\tau_y k_x)$ \\
 $S^y$  & $\tau_y /2$ &  $-\kappa (k_x^2-k_y^2) \tau_y /2$ & $\kappa (k_x^2-k_y^2)\tau_y /2$\\
$\Lambda$ & $-/+$ & $+/-$ & $-/+$ \\
 $\sigma^{SGE}$  & $(-e)N_0\tau (-\alpha)$ & $(-e)N_0\tau (\beta p_F^2)( \kappa p_F^2)\frac{5}{8}$ & $(-e)N_0\tau (-\alpha )( \kappa p_F^2)$\\
 $D^{SGE}$ &  $(-e)N_0\pi  \left(-\alpha  \right)\frac{1}{2}$  & $(-e)N_0\pi  (\beta p_F^2) (\kappa p_F^2) \frac{1}{4}  $  &  $(-e)N_0\pi  \alpha (\kappa p_F^2) \frac{1}{4} $ \\
\end{TAB}
%}
\caption{Effective Hamiltonian for the three pairs of  bands close to the $\Gamma$ point, together with the form of the spin 
polarization operator $S_y$ and the resulting helicities $\Lambda$.
In the absence of disorder, when the scattering time $\tau\rightarrow \infty$, the SGE response has a Drude-like peak with 
weight $D^{SGE}$ as for the longitudinal conductivity.
In the presence of disorder, the Drude-like peak evolves in the DC limit value of the SGE coefficient $\sigma^{SGE}$.}\label{tab1}
\end{table*}
%\endgroup

%The results shown in the fifth line of Tab.\,\ref{tab1} may be modified by two main effects. First, when we move away from 
%the $\Gamma$ point and the energy separation between the pairs of bands is no longer large enough, inter-band effects may 
%become important. Secondly, the inclusion of disorder scattering, responsible for the presence of a finite momentum 
%relaxation time $\tau$ requires the consideration of the so-called vertex corrections, which may change the sign of the SGE.
%In fact, a full disorder analytical calculation, whose details will be given elsewhere, can be carried out in the vicinity of 
%the $\Gamma$ point by using standard field-theory impurity technique. The $6$th row of Tab.\,\ref{tab1} shows how disorder 
%affects the clean limit result. By comparing the $5$th and $6$th row of Tab.\,\ref{tab1}, one sees the for
%the pairs of bands (1,2) and (3,4) there is no change of sign in going from the clean to the dirty case and a sign change is 
%expected at the Lifshitz transition, in agreement with experimental results \cite{Lesne2016}. However,
%for bands (5,6) one finds a sign change between the clean and disordered case and the apparent question arises how the total 
%SGE derives from all three bands. 

In the remaining of this paper, we are going to discuss these important
effects by a means of exact numerical evaluation of the Kubo formula for the
conductivity $\chi^{SGE}(\omega)=\langle\langle J_x ; S^y\rangle\rangle_\omega/[i(\omega+i\eta)]$,
with
\[
\langle\langle J_x ; S^y\rangle\rangle_\omega=\frac{1}{N}\sum_{k,p}(f_p-f_k)
\frac{\langle p|S^y|k\rangle\langle k|J_x|p \rangle}{\omega+i\eta+E_p-E_k},
\]
where $J_x$ is the total charge current (obtained from the usual Peierls
substitution), $S^y$ denotes the total $y$-polarized spin, $E_p$ are the eigenvalues of the Hamiltonian, and $f_p$
is the Fermi function evaluated at $E_p-\mu$. The real part has a Drude-like and a regular contribution, 
\[
   \sigma^{SGE}(\omega)=D^{SGE}\delta(\omega) + \Im\frac{
   \langle\langle J_x ; S^y\rangle\rangle_\omega}{\omega},
\]
with $D^{SGE}=-\pi \Re\langle\langle J_x ; S^y\rangle\rangle_\omega$ which is finite in a clean system, but is expected to
vanish in the presence of disorder, opposite to the behavior of the regular part. The latter comes
without a ``diamagnetic'' contribution (in contrast to  the case of the optical conductivity) which implies the
sum rule $\int\!d\omega\,\sigma^{SGE}(\omega)=0$ for the SGE response function.

\begin{figure}[ttt!]
      \includegraphics[width=8.5cm,clip=true]{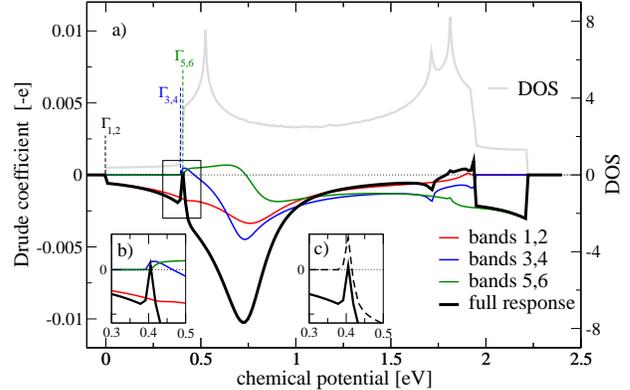}
       \caption{a) SGE Drude coefficient at $T=10$\,K (black)
         and the intraband (red, blue, green) 
         contribution. The DOS is shown in grey for comparison.
         b)  Detail of the intraband contrinution 
         around the Lifshitz point. c) Interband contribution (dashed)
         in the same region as b)
         arising from scattering between bands (3,4) and bands (5,6).
         Calculations have been done for a lattice with $6354\times 6354$ $k$ points.}
\label{fig2}
\end{figure}
   
Fig.\,\ref{fig2} shows the Drude coefficient $D^{SGE}$ for the clean system and its decomposition into the contribution from 
the individual pair of bands as a function of chemical potential $\mu$. Close to the corresponding $\Gamma$ points
(indicated by the vertical dashed lines) the individual contributions
can be evaluated analytically and are given in the $6$th row of 
Tab.\,\ref{tab1} and the $\mu$ dependence is implicit in the corresponding DOS $N_0(\mu)$. Up to $\mu \approx 0.4$\,eV the full 
Drude response (black) is well described by the contribution of the lowest pair of bands (1,2) because the inter-band
contributions arising from the mixing $(1,2)\leftrightarrow (3,4)$ and $(1,2)\leftrightarrow (5,6)$ are opposite in sign and almost compensate.
The inter-band scattering $(3,4)\leftrightarrow (5,6)$ becomes significant upon crossing the Lifshitz point at 
$\mu\approx 0.4$\,eV [cf. panel c)]. In this regime the small positive
contributions from bands (3,4) and (5,6) [cf. 6th row of Tab. \ref{tab1}]
are overcompensated by the negative $D^{SGE}$ from bands (1,2)
together with a large negative inter-band contribution
$(3,4)\leftrightarrow (5,6)$ which 
prevents the occurrence of a significant sign change in $D^{SGE}$ as a function
of $\mu$. This sign change around the Lifshitz 
point is therefore very weak and confined to a very narrow chemical potential
range (cf. panel b) of Fig.\,\ref{fig2}). Another sign change occurs in the high-filling regime around $\mu\approx 1.7$\,eV.

\begin{figure}[ttt!]
      \includegraphics[width=8.5cm,clip=true]{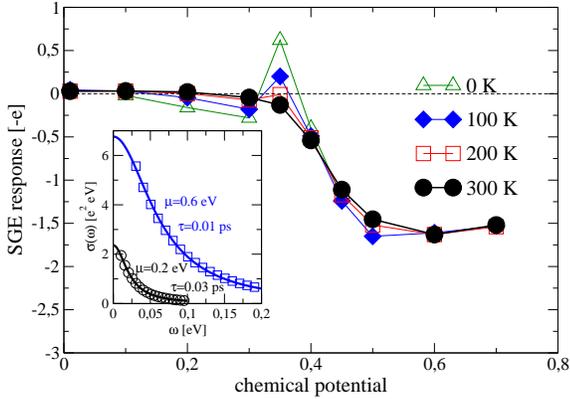}
      \caption{Regular part of the SGE response function as a
        function of chemical potential. Disorder potential
        is $V_0=0.1$\,eV and calculations are performed on $24\times 24$ lattices.}
\label{fig3}
\end{figure}

The Drude SGE response is obtained in the clean limit which corresponds to a
non-stationary situation when there is no dissipation and the momentum relaxation time $\tau$ becomes infinite. However, Fig. \ref{fig2}
suggests that the experimentally observed sign change in $\sigma^{SGE}$
around the Lifshitz point \cite{Lesne2016} may be supported by disorder
which changes the relative contributions of intra- and inter-band processes to
the SGE. We therefore proceed by
evaluating the response for the disordered system by adding an impurity potential
$V=\sum_{n} V_n {\bf 1}$ in the local basis of the $t_{2g}$ orbitals to the Hamiltonian Eq.\,(\ref{hamiltonian}). 
The $V_n$ are drawn from a flat distribution $-V_0\le V_n \le V_0$ and we average over typically $50$ impurity 
realizations and $100$ boundary conditions in order to minimize finite size effects.
For a disorder potential $V=0.2$\,eV the inset to Fig.\,\ref{fig3}
shows the optical conductivity for two values of the chemical
potential below and above the Lifshitz point. By fitting with the Drude formula we find a transport scattering time 
$\tau \approx 0.3$\,ps for $\mu=0.2$\,eV and $\tau \approx 0.01$\,ps for $\mu =0.6$\,eV,
which is of the same order of magnitude as that obtained with magnetotransport measurements on LAO/STO 
heterostructures \cite{liang15}. The SGE response in the presence of disorder is shown in the main panel of Fig.\,\ref{fig3}. 
Compared to the clean case one observes a significant suppression in the regime where the chemical
potential falls within the $d_{xy}$-bands. The sign change around the Lifshitz point is still confined to a narrow window, but 
is more pronounced than in the clean case. Finally, when the chemical potential enters the $d_{xz}\/d_{yz}$-bands we 
obtain a sizable SGE response and the overall behavior is in very good agreement with the gate voltage dependence of the 
spin galvanic effect as measured in Ref.\,\cite{Song2017} at $T=300$\,K. In contrast, the voltage dependence of the 
SGE response in Ref.\,\cite{Lesne2016} has been measured at much lower temperature $T=7$\,K and shows the sign change 
upon voltage reversal. At low temperature our results
  display two sign changes: from negative to positive $\sigma^{SGE} [-e]$
  when $\mu$ is enters into bands (3,4) and back to negative $\sigma^{SGE} [-e]$
  when $\mu$ moves inside bands (5,6). The latter we therefore associate with
  the experimentally observed sign change and one may
expect a second sign reversal of the SGE when experimentally the gating could be
tuned to larger negative values at low temperature, in
agreement with our analytical analysis which in the diffusive limit
yields a {\it negative} 
$\sigma^{SGE} [-e]$ close to the bottom of the xy-type bands (1,2)
[cf. Tab. \ref{tab1}]. In this regard one may also speculate whether a more 
realistic implementation of disorder may help to extend the region
of positive $\sigma^{SGE} [-e]$ deeper into the $d_{xy}$-band range (i.e. 
toward negative gate voltages). In fact, since $d_{xz}$- and $d_{yz}$-orbitals extend deeper into the STO bulk it is plausible 
to assume that they are less affected by disorder in the 2DEG system than the $d_{xy}$-states. This is also supported by the 
gate voltage dependence of the transport scattering time which shows a pronounced increase upon crossing the Lifshitz 
point \cite{liang15} and a refinement of our computations in this regard opens a perspective for future investigations. Moreover, 
the Rashba SOC in STO-based heterostructures has been recently also
analyzed within a ${\bf k}\cdot {\bf p}$ Luttinger-Kohn (LK) approach \cite{heeringen13,heeringen17}
which provides an alternative to the asymmetry term $H_I$ in our hamiltonian.
Differences occur in particular around the Lifshitz point where the LK model
predicts a smaller Rashba SOC for the lowest pair of bands. Future studies
should therefore be directed at clarifying the consequences of the LK model
on the SGE and other transport response functions.

\bibliography{paper_bib}

\begin{thebibliography}{27}
\bibitem [{not()}]{note1}%
  \BibitemOpen
  \href@noop {} {}\bibinfo {note} {Note that for the splitting $\Delta$ we take
  a value between the theoretical ($\Delta=0.19$\,eV) end the experimental one
  ($\Delta=0.6$\,eV). As long as $\Delta\gg \Delta_{aso}$ and $\Delta \gg
  \gamma$ this has no influence on our further results.}\BibitemShut {Stop}%
\end{thebibliography}

%{prsty} 
%\bibliographystyle{abbrv}
\end{document}